\documentclass [leqno,11pt] {article}
\usepackage{latexsym}
\usepackage{epsfig}
\usepackage{color}
\newcommand{\N}{{I\!\!N}}
\newcommand{\Z}{{Z\!\!\!Z}}
\newcommand{\hoare}[3]{\{#1\}~#2~\{#3\}}
\newcommand{\regle}[2]{{\Large $\frac{#1}{#2}$ \normalsize}}
\newcommand{\msubst}[3]{{#1 [ #2\leftarrow#3 ]}}
\newcommand{\cqfd}{\hfill$\Box$}
\newcommand{\relation}[1]{\stackrel{#1}{\leadsto}}
\newtheorem{theorem}{Theorem}[section]

\newtheorem{corollary}[theorem]{Corollary}
\newtheorem{notation}[theorem]{Notation}
\newtheorem{definition}[theorem]{Definition}
\newtheorem{remark}[theorem]{Remark}

\begin{document}
\newpage\begin{center}{\Large\bfseries
	A Genetically Modified Hoare Logic 
}\end{center}
\begin{center}
G. Bernot$^1$, 
J.-P. Comet$^1$, 
Z. Khalis$^1$, 
A. Richard$^1$, 
O. Roux$^2$

$^1$ University Nice-Sophia Antipolis\\
I3S laboratory, UMR CNRS 7271,  Les Algorithmes, b\^at. Euclide B, BP.121,\\
06903 Sophia Antipolis Cedex, France

$^2$ IRCCyN UMR CNRS 6597, BP 92101, \\
1 rue de la No\"e, 44321 Nantes Cedex 3, France
\end{center}
\begin{quote}
\textbf{Abstract:}
An important problem when modeling gene networks lies in the
identification of parameters, even if we consider a purely discrete
framework as the one of Ren\'e Thomas.  Here we are interested in the
exhaustive search of all parameter values that are consistent with
observed behaviors of the gene network.  We present in this article a
new approach based on Hoare Logic and on a weakest precondition
calculus to generate constraints on possible parameter
values. Observed behaviors play the role of ``programs'' for the
classical Hoare logic, and computed weakest preconditions represent
the sets of all compatible parameterizations expressed as constraints
on parameters.
Finally we give a proof of correctness of our Hoare logic
for gene networks as well as a proof of completeness based on the
computation of the weakest precondition.
\end{quote}

\vspace{-\parskip}\section{
	Introduction
}\label{intro}\vspace{-\parskip}

Gene regulation is a complex process where the expression level of a
gene at each time depends on a large amount of interactions with
related genes. Hence regulations between genes can be seen as a
gene network.  Different methods for studying the behavior of
such gene networks in a systematic way have been
proposed. Among them, ordinary differential equations played an
important role which however mostly lead to numerical
simulations. Moreover, the nonlinear nature of gene regulations makes
analytic solutions hard to obtain. Besides, the abstraction procedure
of Ren\'e Thomas~\cite{TH091}, approximating sigmoid functions by step
functions, makes it possible to describe the qualitative dynamics 
of gene networks as paths in a finite state space.
Nevertheless this qualitative description of the dynamics is governed
by a set of parameters which remain difficult to
be deduced from classical experimental knowledge. Therefore, even
when modeling with the discrete approach of Ren\'e Thomas, the main
difficulty lies in the
identification of these parameters.  In this context, we are interested
in the exhaustive search of parameter values which are consistent with
specifications given by the observed behavior of gene regulatory
networks. Because of the exponential number of parameterizations to
consider, two main kinds of approaches have emerged. On the one hand,
information about cooperation or concurrence between two regulators of
a same target can be taken into account in order to reduce the number
of parameterizations to consider, see for example~\cite{GGG2009} and
also~\cite{Corblin2009} in which this notion of cooperation is treated
\emph{via} a grouping of states. On the other hand, using constraints can be
helpful to represent the set of consistent parameterizations see for
example~\cite{FanchonCTHG04,Corblin2009,articleJBCB2007}.

In this paper, we present a new approach based on Hoare
Logic~\cite{logique-Hoare-1969} and on weakest precondition
calculus~\cite{Dijkstra:1975:GCN:360933.360975} to generate
constraints on parameters. A feature of this approach lies in the fact
that specifications are partially described by a set of paths, seen as
``programs.''  Since this method avoids building the complete state
graph, it results in a powerful tool to find out the constraints
representing the set of consistent parameterizations with a tangible
gain for CPU time. Indeed, the weakest precondition computation which
builds the constraints, goes through the ``program'' but is
independent of the size of the gene network.


Other works were undertaken with such objectives. The application of temporal
logic to biological regulatory networks was presented in
\cite{articleJTB2004}. Constraint programming was used for biological systems
in \cite{boileau01} and these ideas were continued specifically for genetic
regulatory networks in \cite{CORBLIN:2008:TEL-00388776:1,Corblin2009}.

The paper is organized as follows. The basic concepts of Hoare logic and
Dijkstra weakest precondition are quickly reminded in Section~\ref{stdHoare}.
The formal definitions for discrete gene regulatory networks
are given in Section~\ref{multiplexes}. 
%
Section~\ref{prepost} gives the way to describe properties of states,
then presents the path language, and finally introduces the notion of
Hoare triplet.  The semantics of these extended Hoare triples is given
in Section~\ref{semantics}.
With the previous material, in Section~\ref{discreteHoare} an extended
Hoare logic for gene networks is defined for Thomas' discrete models.
In Section~\ref{example}, the example of the ``incoherent feedforward
loop of type 1'' (made popular by Uri Alon
in~\cite{Alon2002-a,Alon2002-b}) highlights the whole process of our
approach to find out the suitable parameter values.
Section~\ref{WPcorrectness} contains a proof of correctness of our
Hoare logic for gene networks as well as a proof of
completeness based on the computation of the weakest precondition. We
conclude in Section~\ref{discussion}.
\vspace{-\parskip}\section{
	Reminders on standard Hoare logic
}\label{stdHoare}\vspace{-\parskip}
Hoare logic is a formal system for reasoning about the correctness of
imperative programs. In~\cite{logique-Hoare-1969}, \emph{C. A. R. Hoare}
introduced the notation ``$\hoare{P}{pgm}{Q}$'' to mean ``If the
assertion $P$ (precondition) is satisfied before performing the program
$pgm$ and if the program terminates, then the assertion $Q$
(postcondition) will be satisfied afterwards.'' This constitutes \textit{de
facto} a specification of the program under the form of a
triple, called the Hoare triple. 

In~\cite{Dijkstra:1975:GCN:360933.360975}, E. W. Dijkstra has defined an
algorithm taking the postcondition $Q$ and the program $pgm$ as input and
computing the \emph{weakest precondition} $P_0$ that ensures $Q$ if $pgm$
terminates. In other words, the Hoare triple $\hoare{P_0}{pgm}{Q}$ is
satisfied and, for any precondition $P$, $\hoare{P}{pgm}{Q}$ is
satisfied if and only if $P\Rightarrow P_0$.

Hoare logic and weakest preconditions are now widely known and teached
all over the world. The basic idea is to stamp the sequential phases
of a program with assertions that are infered according to the
instruction they surround. There are several equivalent versions of
Hoare logic and our prefered one is the following because it offers a
simple proof strategy to compute the weakest precondition \textit{via}
a proof tree. Here, $p$, $p_1$ and $p_2$ stand for programs, $P$,
$P_1$, $P_2$, $I$ and $Q$ stand for assertions, $v$ stands for a
declared variable of the imperative program, and $\msubst{Q}{v}{expr}$
means that $expr$ is substituted to each free occurrence of $v$ in
$Q$.

\begin{tabular}{ll}
\textbf{Assignment rule:} &
	\regle{}{\hoare{\msubst{Q}{v}{expr}}{v \mathtt{:=} expr}{Q}}
\\[3ex]
\textbf{Sequential composition rule:} &
	\regle{\hoare{P_{2}}{p_{2}}{Q}~~~~~\hoare{P_{1}}{p_{1}}{P_{2}}}
	{\hoare{P_{1}}{p_{1};p_{2}}{Q}}
\\[3ex]
\textbf{Alternative rule:} &
	\regle{\hoare{P_{1}}{p_{1}}{Q}~~~~~\hoare{P_{2}}{p_{2}}{Q}}
	{\hoare{(e\wedge P_{1})\vee (\neg e\wedge P_{2})}
		{\mathtt{if}~e~\mathtt{then}~p_{1}~\mathtt{else}~p_{2}}
		{Q}}
\\[3ex]
\textbf{Iteration rule:} &
	\regle{\hoare{e\wedge I}{p}{I}}
	{\hoare{I}
		{\mathtt{while}~e~\mathtt{with}~I~\mathtt{do}~p}
		{\neg e\wedge I}}
\\[3ex]
\textbf{Empty program rule:} &
	\regle{P \Rightarrow Q}
	{\hoare{P}{\varepsilon}{Q}}
        \quad(where $\varepsilon$ stands for the empty program)
\end{tabular}

There are standard additional rules: first order logic (to establish
``$P\Rightarrow Q$'' introduced by the Empty program rule), and in practice
some reasonnings about the data structures of the program (e.g. integers) in
order to simplify the expressions as much as possible  ``on the fly'' in the
proof trees.

The \emph{Iteration rule} requires some comments. The assertion $I$ is called the
\emph{loop invariant} and it is well known that finding the weakest loop
invariant is undecidable~\cite{HATCHER1974,Blass2001}. 
It has been included within the programming language
for this reason; we ask the programmer to give a loop invariant explicitely
after the \texttt{with} keyword, although it may appear redundant as it is
also the precondition of the Hoare triple. By doing so, within a program, each
\texttt{while} instruction carries its own (sub)specification and it can
consequently be proved apart from the rest of the program.

When using these Hoare logic rules, the following proof strategy builds a
proof tree that performs the proof by computing the weakest
precondition~\cite{Dijkstra:1975:GCN:360933.360975}:

\begin{enumerate}
	\item
For each \texttt{while} statement within a Hoare triple 
\begin{center}
$\hoare{P}{p_1 ~;~ \mathtt{while}~e~\mathtt{with}~I~\mathtt{do}~p_2 ~;~
p_3}{Q}$
\end{center}
perform 3 independent sub-proofs:
	\begin{itemize}
		\item
	$\hoare{\neg e\wedge I}{p_3}{Q}$
		\item
	$\hoare{I} {\mathtt{while}~e~\mathtt{with}~I~\mathtt{do}~p_2} {\neg
	e\wedge I}$ ~(i.e.~$\hoare{e\wedge I}{p_2}{I}$ ~according to the
	Iteration rule)
		\item
	$\hoare{P}{p_1}{I}$
	\end{itemize}
This first step of strategy leads to proofs on subprograms that do
not contain any \texttt{while} instruction.
	\item
Apply the \emph{Sequential composition rule} only when the program $p_2$ of this
rule is reduced to an instruction, which leads to perform the proof 
starting from postcondition $Q$ at the end and treat the instructions backward.
	\item
Never apply the \emph{Empty program rule}, except when the leftmost
instruction has been treated, that is when all instructions have been treated.
\end{enumerate}

Since the \emph{Assignment rule}, which is central, makes it possible to
precisely define \emph{one} precondition from its postcondition, and since the
other rules relate (but do not evaluate) the conditions, the proof tree is
done from the end to the beginning of the program and computes a unique
assertion. By doing so, the precondition obtained just before applying the
last Empty program rule is actually the weakest precondition (assuming that
the programmer has given the weakest loop invariants). In the remainder of the
article, we call this strategy \emph{the backward strategy}. In
Section~\ref{example}, we always follow the backward strategy.

The most striking feature of Hoare logic and weakest precondition is that
proofs according to the backward strategy consist in very simple sequences of
syntactic formula substitutions and end with first order logic proofs.
Nevertheless, it is worth noticing that it is only a question of
\emph{partial} correctness since Hoare Logic does not give any proof of the
termination of the analyzed program (\texttt{while} instructions may induce
infinite loops).

\vspace{-\parskip}\section{
Discrete gene regulatory networks with multiplexes
}\vspace{-\parskip}\label{multiplexes}

This section presents our modeling framework, based on the general
discrete method of Ren\'e Thomas {\cite{TA90,TK2001-b}} and introduced
in {\cite{GGG2009,ZohraKhalisPhD}}. 

The starting point consists in a labeled directed graph in which
vertices are either {\emph{variables}} or
{\emph{multiplexes}}. Variables abstract genes or their products, and
multiplexes contain propositional formulas that encode situations in
which a group of variables (inputs of multiplexes) influence the
evolution of some variables (outputs of multiplexes). Hence
multiplexes represent biological phenomena, such as the formation
of complexes to activate some genes. In the next definition, this
labeled directed graph is formally defined, and it is associated with
a family $\mathcal{K}$ of integers. As we will see later, these
integers correspond to parameters that drive the dynamics of the network.

\begin{definition}
A \emph{gene regulatory network with multiplexes} (\textsc{grn} for short) is
a tuple $N=(V,M,E_{V},E_{M},\mathcal{K})$ satisfying the following conditions:
\begin{itemize}
\item
$V$ and $M$ are disjoint sets,
whose elements are called {\emph{variables}} and {\emph{multiplexes}}
respectively.
\item
$\tilde N= (V\cup M,E_{V}\cup E_{M})$ is a labeled
directed graph such that:
\begin{itemize}
\item
edges of $E_{V}$ start from a variable and end to a multiplex, and
edges of $E_{M}$ start from a multiplex and end to either a
variable or a multiplex.
\item
every directed cycle of $\tilde N$ contains at least one variable.
\item
every variable $v$ of $V$ is labeled by a positive integer $b_{v}$
called the \emph{bound} of $v$.
\item
every multiplex $m$ of $M$ is labeled by a formula $\varphi _{m}$
belonging to the language~$\mathcal{L}_{m}$ inductively defined by:
\begin{enumerate}
\item[$-$]
If $v\rightarrow m$ belongs to $E_V$ and $s\in\N$ belongs to the interval
$[1,b_v]$, then $v\geq s$ is an atom of $\mathcal{L}_{m}$.
\item[$-$]
If $m'\rightarrow m$ belongs to $E_{M}$ then $m'$ is an atom of
$\mathcal{L}_{m}$.
\item[$-$]
If $\varphi$ and $\psi$ belong to $\mathcal{L}_{m}$ then $\neg \varphi
$, $(\varphi \wedge \psi )$ and $(\varphi \vee \psi )$ also belong to
$\mathcal{L}_{m}$.
\end{enumerate}
\end{itemize}
\item
$\mathcal{K} = \{ K_{v,\omega }\}$ is a family of integers indexed by $v\in V$
and $\omega \subset N^{-1}(v)$, where $N^{-1}(v)$ is the set of predecessors
of $v$ in $\tilde N$ (that is, the set of multiplexes $m$ such that $m\to v$ is
an edge of $E_M$). Each $K_{v,\omega }$ must satisfy $0\leq K_{v,\omega }\leq
b_{v}$.
\end{itemize}
\end{definition}

\begin{notation}
The \emph{flaten version} of a formula $\varphi _{m}$,  denoted
$\overline{\varphi _{m}}$, is obtained by applying the following
algorithm: while the formula contains a multiplex atom $m'$, substitute $m'$
by its associated formula $\varphi _{m'}$. The formula $\overline{\varphi
_{m}}$ exists since $\tilde N$ has no directed cycle with only multiplexes. As
a result, all the atoms of $\overline{\varphi _{m}}$ are of the form $v\geq
s$.
\end{notation}

A {\emph{state}} is an assignment of integer values to the variables of $V$.
Such an assignment allows a natural evaluation of any formula $\varphi_m$: by
replacing variables by their values, $\overline{\varphi_m}$ becomes a
propositional formula whose atoms are integer inequalities.

\begin{definition}\label{DEFresources}\emph{(states, satisfaction
relation and resources)}.
Let $N$ be a \textsc{grn} and $V$ be its set of variables. A {\emph{state}} of
$N$ is a function $\eta : V \rightarrow \N $ such that $\eta (v)\leq b_{v}$
for all $v\in V$. Let $\mathcal{L}$ be the set of propositional formula whose
atoms are of the form $v\geq s$ with $v\in V$ and $s$ be a positive integer (so
that $\overline{\varphi_m}$ is a formula of $\mathcal{L}$ for every multiplex
$m$ of $N$). The satisfaction relation $\models_N$ between a state $\eta$ of
$N$ and a formula $\varphi$ of $\mathcal{L}$ is inductively defined~by:
\begin{itemize}
\item
if $\varphi$ is reduced to an atom of the form $v\geq s$, then
$\eta\models_N \varphi$ if and only if $\eta(v)\geq s$.
\item
if $\varphi\equiv \psi_1\land \psi_2$ then $\eta\models_N \varphi$ if and only if
$\eta\models_N \psi_1$ and $\eta\models_N \psi_2$; and we proceed similarly for
the other connectives.
\end{itemize}
Given a variable $v\in V$, a multiplex $m\in N^{-1}(v)$ is a {\emph{resource}}
of $v$ at state $\eta$ if $\eta\models_N \overline{\varphi _{m}}$. The
\emph{set of resources} of $v$ at state $\eta$ is defined by
$\rho(\eta,v)=\{m\in N^{-1}(v)~|~\eta \models _{N}\overline{\varphi _{m}}\}$.
\end{definition}

From a dynamical point of view,
at a given state $\eta$, each variable $v$ is supposed to evolve in
the direction of a specific level (between $0$ and $b_v$) that only
depends on the set $\rho(\eta,v)$. This {\emph{focal level}} is given
by the {\emph{logical parameter}} $K_{v,\rho(\eta,v)}$. Hence, at
state $\eta$, $v$ can increase if $\eta(v)<K_{v,\rho(\eta,v)}$, it can
decrease if $\eta(v)>K_{v,\rho(\eta,v)}$, and it is stable if
$\eta(v)=K_{v,\rho(\eta,v)}$.

Suppose for instance that $v$ has two input multiplexes $m_{ab}$ and $m_{cd}$
with formula $(a\geq 1\land b\geq 1)$ and $(c\geq 1\land d\geq 1)$ respectively.
Then $m_{ab}$ and $m_{cd}$ may be seen as complexes (dimers) regulating the
level of $v$. Suppose, in addition, that $K_{v,\emptyset}=0$,
$K_{v,\{m_{ab}\}}=K_{v,\{m_{cd}\}}=1$ and $K_{v,\{m_{ab},m_{cd}\}}=2$. Then,
complexes $m_{ab}$ and $m_{cd}$ specify activator complexes, with an individual
effect which is less than a cumulated effect (the focal level of $v$ in the
presence of a single complex is less than the focal level of $v$ in the
presence of both complexes). This example illustrates the fact that
multiplexes encode combinations of variables that regulate a given variable,
and that the parameters, by giving a weight to each possible
combinations of multiplexes, indicate how multiplexes regulate
a given variable.

As in Thomas' method {\cite{TA90,TK2001-b}}, it is assumed that variables
evolve asynchronously and by unit steps toward their respective target levels.
The dynamics of a gene regulatory network is then described by the
following asynchronous state graph.

\begin{definition}\label{StateGraph} \emph{(State Graph)}.
Let $N=(V,M,E_V,E_M,\mathcal{K})$ be a \textsc{grn}. The
\emph{state graph} of $N$ is the directed graph
$\mathcal{S}$ defined as follows: the set of vertices is the set of states of
$N$, and there exists an edge (or transition) $\eta\to\eta'$ if one of
the following conditions is satisfied:
\begin{itemize}
\item 
there is no $v\in V$ such that $\eta(v)\neq K_{v,\rho(\eta,v)}$ and $\eta'=\eta$.
\item
there exists $v\in V$ such that $\eta(v)\neq K_{v,\rho(\eta,v)}$ and 
\[
\eta'(v)=
\left\{
\begin{array}{ll}
\eta(v)+1&\textrm{if }\eta(v)< K_{v,\rho(\eta,v)}\\
\eta(v)-1&\textrm{if }\eta(v)> K_{v,\rho(\eta,v)}
\end{array}
\right.
\quad \textrm{ and } \quad
\forall u\neq v, ~ \eta'(u)=\eta(u).
\]
\end{itemize}
\end{definition}

Hence a state $\eta$ is a stable state if and only if it has itself as
successor, that is, if and only if every variable is stable at state
$\eta$ ({\emph{i.e.}} $\eta(v)= K_{v,\rho(\eta,v)}$ for every variable
$v$). If $\eta$ is not a stable state, then it has at least one
outgoing transition. More precisely, for
each variable $v$ such that $\eta(v)\neq K_{v,\rho(\eta,v)}$, there is
a transition allowing $v$ to evolve ($\pm 1$) toward its focal level
$K_{v,\rho(\eta,v)}$. Every
outgoing transition of $\eta$ is supposed to be possible, so that
there is an indeterminism as soon as $\eta$ has several outgoing
transitions. An example is given in Figure~\ref{figure:example1} (see also
Section~\ref{section:example} for another example).
\vspace{-\parskip}\section{
	Pre- and post-conditions on path sets
}\vspace{-\parskip}\label{prepost}

In order to formalize known information about a gene network, we
introduce in this section a language to express properties of states
(assertion language) and a language to express properties of state
transitions (path language). Combining
properties of state transitions and properties of states, at the
beginning and at the end of a sequences of state
transitions, leads to the notion of Hoare triplet on path programs.

\subsection{
	An assertion language for discrete models of gene networks
}
To describe properties of states in a meaningful way, we
need terms that allow us to check, compare and manipulate variable values
while taking parameter values into account. The following definitions define a
language suitable for such needs. It extends~\cite{ZohraKhalisPhD}. 

\begin{definition}\emph{(Terms of the assertion language)}
Let $N = (V,M,E_{V},E_{M},\mathcal{K})$ be a \textsc{grn}. The \emph{well
formed terms of the assertion language} of $N$ are inductively defined by:
\begin{itemize}\vspace{\itemsep}
	\item\vspace{-\itemsep}
Each integer $n\in \N$ constitutes a well formed constant term
	\item\vspace{-\itemsep}
For each variable $v\in V$, the name of the variable $v$, considered as a
symbol, constitutes a well formed constant term.
	\item\vspace{-\itemsep}
Similarly, for each $v\in V$ and for each subset $\omega$ of $N^{-1}(v)$, the
symbol $K_{v,\omega}$ constitutes a well formed constant term.
	\item\vspace{-\itemsep}
If $t$ and $t'$ are well formed terms then $(t+t')$ and $(t-t')$ are also well
formed terms.
\end{itemize}
\end{definition}

\begin{definition}\emph{(Assertion language and its semantics)}
Let $N = (V,M,E_{V},E_{M},\mathcal{K})$ be a \textsc{grn}. The \emph{assertion
language} of $N$ is inductively defined as follows:
\begin{itemize}\vspace{\itemsep}
	\item\vspace{-\itemsep}
If $t$ and $t'$ are well formed terms then $(t=t')$, $(t < t')$, $(t > t')$,
$(t\leq t')$ and $(t\geq t')$ are atoms of the assertion language.
	\item\vspace{-\itemsep}
If $\varphi$ and $\psi$ belong to the assertion language then $\neg \varphi$,
$(\varphi \wedge \psi)$, $(\varphi \vee \psi)$ and $(\varphi \Rightarrow
\psi)$ also belong to the assertion language.
\end{itemize}

A state $\eta$ of the network $N$ \emph{satisfies} an assertion $\varphi$ if
and only if its interpretation is valid in $\Z$, after substituting each
variable $v$ by $\eta(v)$ and each symbol $K_{v,\omega}$ by its value
according to the family $\mathcal{K}$. We note $\eta \models_{N} \varphi$.
\end{definition}

\subsection{
        A path language for discrete models of gene networks
}
The assertion language introduced above is a subset of first order
logic well suited to describe properties on sets of states. It does
not express dynamical aspects, since the dynamics of the system is
encoded in the transitions of the state graph. A description of
dynamical properties equates to a precise formulation of properties of
paths. The language proposed here is suitable for
encoding such properties.

\begin{definition}\label{PATHlanguage}
\emph{(Path language and path program)}
Let $N = (V,M,E_{V},E_{M},\mathcal{K})$ be a \textsc{grn}. The \emph{path
language} of $N$ is the language inductively defined by:
\begin{itemize}\vspace{\itemsep}
	\item\vspace{-\itemsep}
For each $v\in V$ and $n\in\N$ the expressions ``$v+$'', ``$v-$'' and
``$v:=n$'' belong to the path language (respectively increase, decrease or
assignment of variable value).
	\item\vspace{-\itemsep}
If $e$ is a formula belonging to the assertion language of $N$, then 
``$assert(e)$'' also belongs to the path language. 
	\item\vspace{-\itemsep}
If $p_{1}$ and $p_{2}$ belong to the path language then $(p_{1};p_{2})$
also belongs to the path language (sequential composition). Moreover the sequential
composition is associative, so that we can write $(p_{1};p_{2};\cdots ;p_{n})$
without intermediate parentheses.
	\item\vspace{-\itemsep}
If $p_{1}$ and $p_{2}$ belong to the path language and if $e$ is a
formula belonging to the assertion language of $N$, then
$(if~e~then~p_{1}~else~p_{2})$ also belongs to the path language.
	\item\vspace{-\itemsep}
If $p$ belongs to the path language and if $e$ and $I$ are
formulas belonging to the assertion language of $N$, then
$(while~e~with~I~do~p)$ also belongs to the path language. The assertion
$I$ is called the invariant of the $while$ loop.
	\item\vspace{-\itemsep}
If $p_{1}$ and $p_{2}$ belong to the path language then $\forall(p_{1},p_{2})$
and $\exists(p_{1},p_{2})$ also belong to the path language
(quantifiers). Moreover
the quantifiers are associative and commutative, so that we can write
$\forall(p_{1},p_{2},\cdots,p_{n})$ and $\exists(p_{1},p_{2},\cdots,p_{n})$ as
useful abbreviations.
\end{itemize}
For technical purposes, we also consider the \emph{empty program}
``$\varepsilon$'' (outside the inductive definition).
A well formed expression in the path language is called a \emph{path program}.
\end{definition}


Intuitivelly, ``$v+$'' (resp. ``$v-$'', ``$v:=n$'') means that the
level of variable $v$ is increasing by one unit (resp. decreasing by
one unit, set to a particular value $n$). ``$assert(e)$'' allows one
to express a property of the current state without change of
state. The sequential composition allows one to concatenate two path
programs whereas the statement ``$if$'' allows one to choose between
two programs according to the evaluation of the formula $e$. Finally
it becomes possible to express properties of several paths thanks to
the quantifiers $\forall$ and $\exists$. Lastly notice that
$\varepsilon$ appears in a path program if and only if the path is
reduced to the empty program.  These intuitions will be formalized in
Section~\ref{semantics}.

\subsection{
	Syntax of pre- and post-conditions on path programs
}

The next step is to combine properties of state transitions (path
program) and properties of states (assertions), at the begining and at
the end of the considered path program. This is done \emph{via} the
notion of Hoare triplet on path programs.

\begin{notation}
A \textsc{grn} $N$ being given, a \emph{Hoare triple} on path programs is an
expression of the form ``$\hoare{P}{p}{Q}$'' where $P$ and $Q$ are well formed
assertions, called pre- and post-condition respectively, and $p$ is a
path program. 
\end{notation}
Intuitively, the precondition $P$ describes a set of states, e.g. all states
for which variable $v$ has value zero ($P\equiv v=0$), the path program $p$
describes dynamical processes, e.g. increase of variable $v$ ($p\equiv
v+$), and the postcondition again describes a set of states, e.g. all states
for which variable $v$ has value one ($Q\equiv v=1$). This small example
encodes the process of variable $v$ changing its value from zero to one.
Whether or not the expression is satisfied for a given gene network $N$
depends on its state transition graph, thus it depends on the corresponding parameter
values in $\cal K$.
\vspace{-\parskip}\section{
	Semantics of Hoare triples on path programs
	}\vspace{-\parskip}\label{semantics}

We firstly define the semantics of path programs \textit{via} a binary
relation. The general ideas that motivate the definition below are the
following:
\begin{itemize}
	\item
Starting from an initial state $\eta$, sequences of instructions
without existential or universal quantifier either transform $\eta$
into another state $\eta'$ or is not feasible so that $\eta'$ is
undefined. For example, the simple instruction $v+$ transforms $\eta$ into
$\eta'$ (where $\forall u\neq v, \eta'(u) = \eta(u)$ and $\eta'(v)
= \eta(v)+1$)  if $\eta \rightarrow \eta'$ exists. If, on the
contrary, this transition
does not exist, the instruction is not feasible. 

	\item
Existential quantifiers induce a sort of ``non determinism'' about $\eta'$:
according to the chosen path under each existential quantifier one may get
differents resulting states. Consequently, one cannot define the semantics as
a partial function that associates a unique $\eta'$ to $\eta$; a binary relation
``$\eta\leadsto ...$'' is a more suited mathematical object.
	\item
Universal quantifiers induce a sort of ``solidarity'' between all the states
$\eta'$ that can be obtained according to the chosen path under each universal
quantifier: all of them will have to satisfy the postcondition later on. For
this reason, we define a binary relation that associates a \emph{set of
states} $E$ to the initial state $\eta$: ``$\eta\leadsto E$''. Such a set $E$
can be understood as grouping together the states it contains, under the scope
of some universal quantifier.
	\item
When the path program $p$ contains both existential and universal quantifiers,
we may consequently get several sets $E_1,\cdots,E_n$ such that
``$\eta\relation{p}E_i$'', each of the $E_i$ being a possibility through the
existential quantifiers of $p$ and all the states belonging to a given $E_i$
being together through the universal quantifiers of $p$. On the contrary, if
$p$ is not feasible, then there is no set $E$ at all such that
``$\eta\relation{p}E$''
\end{itemize}

\begin{notation}
For a state $\eta$, a variable $v$ and $k\in[0,b_v]$, 
we define $\msubst{\eta }{v}{k}$ as the state $\eta'$ such that
$\eta'(v) = k$ and for all $u\neq v$, 
$\eta'(u) = \eta(u)$.
\end{notation}

\begin{definition}\label{PATHsemantics}
{\em [Path program relation $\relation{p}$]}. ~~ 

Let $N = (V,M,E_{V},E_{M},\mathcal{K})$ be a \textsc{grn}, let
$\mathcal{S}$ be the state graph of $N$ whose set of vertices is
denoted $S$ and let $p$ be a path program of $N$. 
The binary relation $\relation{p}$ is the smallest subset of
$S\times\mathcal{P}(S)$ such
that, for any state $\eta$:
\begin{enumerate}\vspace{\itemsep}
	\item\vspace{-\itemsep}
If $p$ is reduced to the instruction $v+$ (resp.~$v-$), then let
us consider $\eta ' = \msubst{\eta }{v}{(\eta(v)+1)}$ (resp. $\eta ' =
\msubst{\eta }{v}{(\eta(v)-1)}$): if $\eta \rightarrow \eta '$ is a transition
of $\mathcal{S}$ then $\eta \relation{p} \{\eta '\}$
	\item\vspace{-\itemsep}
If $p$ is reduced to the instruction $v:=k$, then 
$\eta \relation{p} \{\msubst{\eta }{v}{k}\}$ 
	\item\vspace{-\itemsep}
If  $p$ is reduced to the instruction $assert(e)$, if
$\eta \models _{N } e$, then $\eta \relation{p}\{\eta\}$
	\item\vspace{-\itemsep}
If $p$ is of the form $\forall (p_{1},p_{2})$: if $\eta \relation{p_{1}}
E_{1}$ and $\eta \relation{p_{2}} E_{2}$ then $\eta \relation{p} (E_{1}\cup E_{2})$
	\item\vspace{-\itemsep}
If $p$ is of the form $\exists (p_{1},p_{2})$: if $\eta \relation{p_{1}}
E_{1}$ then $\eta \relation{p} E_{1}$, \emph{and} if $\eta \relation{p_{2}} E_{2}$
then $\eta \relation{p} E_{2}$
	\item\vspace{-\itemsep}
If $p$ is of the form $(p_{1};p_{2})$: if $\eta \relation{p_{1}} F$
and if $\{E_{e}\}_{e\in F}$ is a $F$-indexed family of state sets such that 
$e\relation{p_2}E_e$, 
then $\eta \relation{p}(\bigcup_{e \in F}E_{e})$
	\item\vspace{-\itemsep}
If $p$ is of the form $(if~e~then~p_{1}~else~p_{2})$:
	\begin{itemize}\vspace{\itemsep}
		\item\vspace{-\itemsep}
	if $\eta \models _{N } e$ and $\eta \relation{p_{1}} E$ then
	$\eta \relation{p} E$
		\item\vspace{-\itemsep}
	if $\eta \not \models _{N } e$ and $\eta \relation{p_{2}} E$ then
	$\eta \relation{p} E$
	\end{itemize}
	\item\vspace{-\itemsep}
If $p$ is of the form $(while~e~with~I~do~p_{0})$:
	\begin{itemize}\vspace{\itemsep}
		\item\vspace{-\itemsep}
	if $\eta \models _{N } e$ and $\eta \relation{p_{0};p} E$ then
	$\eta \relation{p} E$
		\item\vspace{-\itemsep}
	if $\eta \not \models _{N } e$ then
	$\eta \relation{p} \{\eta \}$
	\end{itemize}
	\item\vspace{-\itemsep}
If $p$ is the empty program $\varepsilon$, then $\eta \relation{p} \{\eta \}$
\end{enumerate}
\end{definition}
This definition calls for several comments.

The relation $\relation{p}$ exists because (i) the set of all relations that
satisfy the properties 1--8 of the definition is not empty (the relation which
links all states to all sets of states satisfies the properties) and (ii) the
intersection of all the relations that satisfy the properties 1--8, also
satisfies the properties.  

A simple instruction such as $v+$ can be not feasible from a state $\eta$ (if
$\eta\rightarrow\eta'$ is not a transition of $\mathcal{S}$). In this case,
there is no set $E$ such that $\eta\relation{v+}E$. The same situation
happens when the program is an assertion that is evaluated to false at
the current state $\eta$.

Universal quantifiers ``propagate'' non feasible paths: if one of the $p_i$ is
not feasible then $\forall(p_1,\cdots,p_n)$ is not feasible. \emph{It is not
the case for existential quantifiers}: if $\eta\relation{p_i}E_i$ for one of
the $p_i$ then $\eta\relation{\exists(p_1\cdots p_n)}E_i$ even if one of the
$p_j$ is not feasible.

When a \emph{while} loop does not terminate, there does not exist a
set $E$ such that $\eta\relation{while...}E$. This is due to the
minimality of the binary relation $\relation{p}$. On the contrary,
when the \emph{while} loop terminates, it is equivalent to a program
containing a finite number of the sub-program $p_0$ in sequence,
starting from $\eta$.

The semantics of sequential composition may seem unclear for whom is not
familiar with commutations of quantifiers. We better take an example to
explain the construction of $\relation{p_1;p_2}$ (see
Figure~\ref{sequenceSemantics}):
\begin{figure}[h]
\centerline{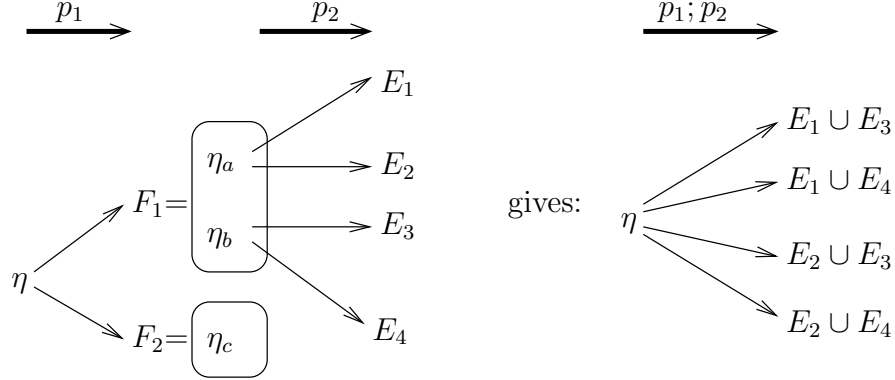}
\caption{An example for the semantics of sequential composition}
\label{sequenceSemantics}
\end{figure}
\begin{itemize}
	\item
Let us assume that starting from state $\eta$, two sets of states are
reachable \textit{via} $p_1$: $\eta\relation{p_1}F_1=\{\eta_a,\eta_b\}$ and
$\eta\relation{p_1}F_2=\{\eta_c\}$. It intuitively means that $p_1$
permits a choice between $F_1$ and $F_2$ through some existential quantifier
on paths and that the path leading to $F_1$ contains a universal quantifier
grouping together $\eta_a$ and $\eta_b$.
	\item
Let us also assume that:
	\begin{itemize}
		\item
	starting from state $\eta_a$, two sets of states are reachable
	\textit{via} $p_2$: $\eta_a\relation{p_2}E_1$ and
	$\eta_a\relation{p_2}E_2$,
		\item
	starting from state $\eta_b$, two sets of states are reachable
	\textit{via} $p_2$: $\eta_b\relation{p_2}E_3$ and
	$\eta_b\relation{p_2}E_4$,
		\item
	there is not any set $E$ such that $\eta_c\relation{p_2}E$
	\end{itemize}
\end{itemize}
When focusing on the paths of $(p_1;p_2)$ that encounter $F_1$ after $p_1$,
the paths such that $p_1$ leads to $\eta_a$ must be grouped together with the
ones that leads to $\eta_b$. Nevertheless, for each of them, $p_2$ permits a
choice: between $E_1$ or $E_2$ for $\eta_a$ and between $E_3$ or $E_4$ for
$\eta_b$. Consequently, when grouping together the possible futures of
$\eta_a$ and $\eta_b$, one needs to consider the four possible combinations:
$\eta\relation{p_1;p_2}(E_1\cup E_3)$, $\eta\relation{p_1;p_2}(E_1\cup E_4)$
$\eta\relation{p_1;p_2}(E_2\cup E_3)$ and $\eta\relation{p_1;p_2}(E_2\cup
E_4)$. \\ Lastly, when focusing on the paths of $(p_1;p_2)$ that encounter
$F_2$ after $p_1$, since $\eta_c$ has no future \textit{via} $p_2$, there is
no family indexed by $F_2$ as mentioned in the definition and consequently it
adds no relation into $\relation{p_1;p_2}$.

Lastly, let us remark that, if $\eta \relation{p} E$ then $E$ cannot be empty;
it always contains at least one state. The proof is easy by structural
induction of the program $p$ (using the fact that a $while$ loop which
terminates is equivalent to a program containing a finite number of the
sub-program~$p_0$). 

\begin{definition}\label{HOAREsemantics}\emph{(Semantics of a Hoare triple)}.
Let $N = (V,M,E_V,E_M,\mathcal{K})$ be a \textsc{grn} and let
$\cal{S}$ be the state graph of $N$ whose set of vertices
is denoted $S$. A Hoare triple
$\hoare{P}{p}{Q}$ is satisfied if and only if: 

for all $\eta\in S$ satisfying $P$, there exists $E$ such that
$\eta\relation{p}E$ 
and for all $\eta'\in E$, $\eta'$ satisfies $Q$.
\end{definition}

The previous definition implies the consistency of all the paths described by
the path program $p$ with the state graph: if path program $p$ is not
feasible from one of the states satisfying pre-condition $P$, then the
Hoare triplet cannot be satisfied. For instance if some $v+$ is
required by the path program $p$ but the increasing of $v$ is not
possible according to the state graph, then the Hoare triple is
not satisfied. 

More generally the path language plays the role of the programming
language in the classical Hoare Logic. A Hoare triplet is
satisfied iff from the states satisfying the precondition, the program
$p$ is feasible and leads to a set of states where the postcondition
is satisfied. The path program $p$ can then be viewed as the sequence
of actions one can use in order to modify the state (memory) of the
system.


Nevertheless, similarly to classical Hoare logic which reflects a
partial correctness of imperative programming language, the previous
definition does not imply termination of $while$ loops. Our path
language can also define some infinite paths. Notice that if the non terminating
$while$ loop is at the end of the program, then it has a biological
meaning: it represents periodic behaviours (such as the circadian cycle
for instance).

\begin{figure}
\centerline{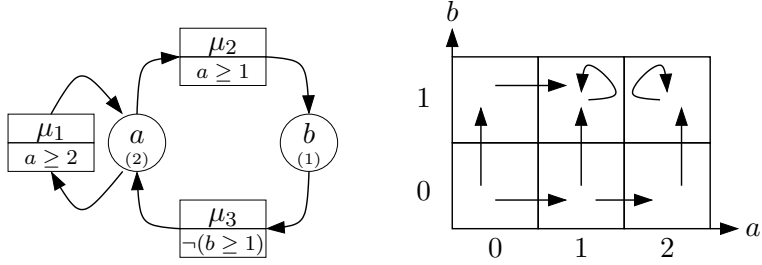}
\caption{{\bf (Left)} Graphical representation of the \textsc{grn}
$N=(V,M,E_V,E_M,{\cal K})$ with $V=\{a,b\}$, the bounds of $a$ 
and $b$ are respectivelly 2 and 1, $M=\{\mu_1,\mu_2,\mu_3\}$,
$\phi_{\mu_1}$ is $(a\geq 2)$, 
$\phi_{\mu_2}$ is $(a\geq 1)$, 
$\phi_{\mu_3}$ is $\lnot(b\geq 1)$. Finally the family of integers is 
$\{K_{a,\emptyset}=1$,  
$K_{a,\{\mu_1\}}=1$,  
$K_{a,\{\mu_3\}}=2$,  
$K_{a,\{\mu_1,\mu_3\}}=2$,  
$K_{b,\emptyset}=1$, 
$K_{b,\{\mu_2\}}=1\}$. {\bf (Right)}
Representation of its state graph.}
\label{figure:example1}
\end{figure}

\paragraph{Examples.}
Let us consider the \textsc{grn} of Figure~\ref{figure:example1} and its state
graph. 
\begin{enumerate}
\item The Hoare triplet $\hoare{(a=0)\land(b=0)}{a+; a+;b+}{(a=2)\land(b=1)}$
      is satisfied, because~: 
      \begin{itemize}
      \item There is a unique state satisfying the precondition
            $(a=0)\land(b=0)$ and 
      \item from this state, the path program $a+; a+; b+$ is possible
      and leads to the state $(2,1)$ and 
      \item the state $(2,1)$ satisfies the postcondition $(a=2)\land(b=1)$.
      \end{itemize}
\item On the opposite, the Hoare triplet $\hoare{(a=2)\land(b=0)}{b+;
a-;a-}{(a=0)\land(b=1)}$ is not satisfied because from the
state satisfying the precondition, the first ``instruction'' $b+$ is
possible and leads to the state $(2,1)$ from which the next
instruction $b-$ is not consistant with the state graph.  

\item The following Hoare triplet contains two existantial quantifiers and
a universal one~:\\
$\hoare{(a=0)\land(b=0)}{\forall(a+,b+); \exists(a+,b+); \exists(\varepsilon,b+)}{(b=1)}$  
      \begin{itemize}
      \item We have clearly 
      $(0,0) \relation{\forall(a+,b+)}\{(1,0),(0,1)\}$ 
      \item Since $(1,0) \relation{\exists(a+,b+)}\{(2,0)\}$ and 
      $(1,0) \relation{\exists(a+,b+)}\{(1,1)\}$ and 
      $(0,1) \relation{\exists(a+,b+)}\{(1,1)\}$, we have both
      $(0,0) \relation{\forall(a+,b+);\exists(a+,b+)}\{(1,1),(2,0)\}$
      and $(0,0) \relation{\forall(a+,b+);\exists(a+,b+)}\{(1,1)\}$. 
      \item We have trivially 
      $(1,1) \relation{\exists(\varepsilon,b+)}\{(1,1)\}$ 
      \item Moreover we have both 
        $(2,0) \relation{\exists(\varepsilon,b+)}\{(2,0)\}$ and 
        $(2,0) \relation{\exists(\varepsilon,b+)}\{(2,1)\}$
      \item We deduce that the considered program $p$ can lead to 3
      differents set of states~: 
      $(0,0) \relation{p}\{(1,1),(2,0)\}$,
      $(0,0) \relation{p}\{(1,1)\}$
      and  
      $(0,0) \relation{p}\{(1,1),(2,1)\}$.  
 
      \end{itemize}
      Because the postcondition is satisfied in both states $(1,1)$
      and $(2,1)$, see the last set of states which is in relation
      with $(0,0)$, one can deduce that the Hoare Triplet is
      satisfied. 
\end{enumerate}

\vspace{-\parskip}\section{
	A Hoare logic for discrete models of gene networks
}\vspace{-\parskip}\label{discreteHoare}
In this section, we define a ``genetically modified'' Hoare logic by giving
the rules for each instruction of our path language
(definition~\ref{PATHlanguage}). First, let us introduce a few notations for
intensively used formulas.

\begin{notation}\label{lesPHI}
Let $N = (V,M,E_{V},E_{M},\mathcal{K})$ be a \textsc{grn} and let $v$ be
a variable of $V$. 
\begin{enumerate}
	\item
For each subset $\omega$ of $N^{-1}(v)$ (set of predecessors of $v$ in the
network), we denote by $\Phi_{v}^{\omega }$ the following formula:
\begin{center}
$\Phi_{v}^{\omega } ~~~ \equiv ~~~ \displaystyle 
(\bigwedge_{m~\in ~\omega }\overline{\varphi _{m}}) ~ \wedge  ~
(\bigwedge_{m~\in ~N^{-1}(v)\setminus \omega }\neg \overline{\varphi
_{m}})$
\end{center}
where $N^{-1}(v)\setminus \omega $ stands for the complementary subset
of $\omega $ in $N^{-1}(v)$.\\
From Definition~\ref{DEFresources}, for all states $\eta$ and for all
variables $v\in V$, $\eta\models_{N}\Phi_{v}^{\omega }$ if and only if
$\omega=\rho(\eta,v)$, that is, $\omega$ is the set of resources of
$v$ at state $\eta$. Consequently, there exists a \emph{unique}
$\omega$ such that $\eta\models_{N}\Phi_{v}^{\omega }$.  

\item We denote by
$\Phi_{v}^{+}$ the following formula:
\begin{center}
$\Phi_{v}^{+} ~~~ \equiv ~~~ \displaystyle 
	\bigwedge_{\omega \subset G^{-1}(v)}
		(\Phi_{v}^{\omega } \Longrightarrow  K_{v,\omega }  >  v)$
\end{center}
From Definition~\ref{StateGraph}, $\eta\models_{N}\Phi_{v}^{+}$ if and only if
there is a transition $(\eta\rightarrow\msubst{\eta }{v}{v+1})$ in the state
graph $\mathcal{S}$, that is, if and only if the variable $v$ can increase.
	\item
We denote by $\Phi_{v}^{-}$ the following formula:
\begin{center}
$\Phi_{v}^{-} ~~~ \equiv ~~~ \displaystyle 
	\bigwedge_{\omega \subset G^{-1}(v)}
		(\Phi_{v}^{\omega } \Longrightarrow  K_{v,\omega }  <  v)$
\end{center}
Similarly, $\eta\models_{N}\Phi_{v}^{-}$ if and only if the variable $v$ can
decrease from the state $\eta$ in the state graph $\mathcal{S}$.
\end{enumerate}
\end{notation}
By the way, in practice, the assertion $assert(\Phi_{v}^{=})$ is often useful
from the biological point of view, where $\Phi_{v}^{=}$ is obviously defined
by: ~~~ $\Phi_{v}^{=} ~~~ \equiv ~~~ \displaystyle 
	\bigwedge_{\omega \subset G^{-1}(v)}
		(\Phi_{v}^{\omega } \Longrightarrow  K_{v,\omega }  =  v)$

Our Hoare logic for discrete models of gene networks is then defined by the
following rules, where $v$ is a variable of the \textsc{grn} and $k\in\N$.
\begin{enumerate}
\item Rules encoding Thomas' discrete dynamics.\\

\begin{tabular}{ll}
\textbf{Incrementation rule:} &
	\regle{}
	{\hoare{~
	\Phi_{v}^{+}~\wedge ~\msubst{Q}{v}{v+1}
	~}{v+}{Q}}
\\[3ex]
\textbf{Decrementation rule:} &
	\regle{}
	{\hoare{~
	\Phi_{v}^{-}~\wedge ~\msubst{Q}{v}{v-1}
	~}{v-}{Q}}
\\[3ex]
\end{tabular}\\
\item Rules coming from Hoare Logic.
These rules are very similar to the ones given in
Section~\ref{stdHoare}. Obvious rules for the instruction $assert(\Phi)$ and the quantifiers
are added:

\noindent
\begin{tabular}{ll}
\textbf{Assert rule:} &
	\regle{}
	{\hoare{~ \Phi ~\wedge ~{Q} ~}{assert(\Phi)}{~Q~}}
\\[3ex]
\textbf{Universal quantifier rule:} &
	\regle{\hoare{P_{1}}{p_{1}}{Q}~~~~~\hoare{P_{2}}{p_{2}}{Q}}
	{\hoare{P_{1}\wedge P_{2}}{\forall (p_{1},p_{2})}{Q}}
\\[3ex]
\textbf{Existential quantifier rule:} &
	\regle{\hoare{P_{1}}{p_{1}}{Q}~~~~~\hoare{P_{2}}{p_{2}}{Q}}
	{\hoare{P_{1}\vee P_{2}}{\exists (p_{1},p_{2})}{Q}}
\\[3ex]
\textbf{Assignment rule:} &
	\regle{}
	{\hoare{
	\msubst{Q}{v}{k}
	}{v:=k}{Q}}
\\[3ex]
\textbf{Sequential composition rule:} &
	\regle{\hoare{P_{2}}{p_{2}}{Q}~~~~~\hoare{P_{1}}{p_{1}}{P_{2}}}
	{\hoare{P_{1}}{p_{1};p_{2}}{Q}}
\\[3ex]
\textbf{Alternative rule:} &
	\regle{\hoare{P_{1}}{p_{1}}{Q}~~~~~\hoare{P_{2}}{p_{2}}{Q}}
	{\hoare{(e\wedge P_{1})\vee (\neg e\wedge P_{2})}
		{if~e~then~p_{1}~else~p_{2}}
		{Q}}
\\[3ex]
\textbf{Iteration rule:} &
	\regle{\hoare{e\wedge I}{p}{I}}
	{\hoare{I}{while~e~with~I~do~p}{\neg e\wedge I}}
\\[3ex]
\textbf{Empty program rule:} &
	\regle{P ~ \Rightarrow ~ Q}
	{\hoare{P}{\varepsilon}{Q}}
\end{tabular}\\
\item Axioms. 
These axioms assert that all values stay between their bounds, where
$v$ is a variable of the \textsc{grn} $N$ and $\omega \subset N^{-1}(v)$:

\noindent
\begin{tabular}{ll}
\textbf{Boundary axioms:} &
	$0\leq v$
\\
~ &
	$v\leq b_{v}$
\\
~ &
	$0 \leq K_{v,\omega }$
\\
~ &
	$K_{v,\omega } \leq b_{v}$
\end{tabular}

\end{enumerate}

\begin{remark}\label{remark:boundaryAxioms}\mbox{}
\begin{itemize}
\item $(\Phi_{v}^{+} \Rightarrow v<b_v)$ can be derived from the
	previous rules. Indeed, $\Phi_{v}^{+}$ implies 
	that for $\omega$ corresponding to the current set of
      resources, $K_{v,\omega} > v$ and, using the boundary axiom
      $K_{v,\omega} \leq b_v$, we get $v<b_v$.
\item Similarly, we have $\Phi_{v}^{-} \Rightarrow v>0$.
\end{itemize}
These implications will be used in section~\ref{example}.
\end{remark}

We will prove in Section~\ref{WPcorrectness} that this modified Hoare logic is
correct, and that it is complete provided that the path program under
consideration contains the weakest loop invariants for all $while$ statements.
More precisely, the proof strategy called \emph{backward strategy}, already
described at the end of Section~\ref{stdHoare}, also applies here: It computes
the weakest precondition. Before giving these two proofs, let us
show in the next section the usefulness of our genetically modified Hoare
logic \textit{via} the formal study of the possible biological functions of a
very simple network.
\vspace{-\parskip}\section{
	Example
}
\begin{figure}[h]
\begin{center}
\hfill
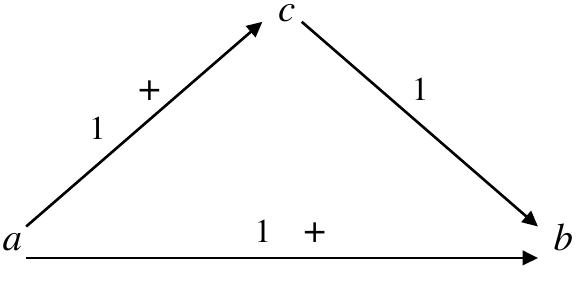
\hfill
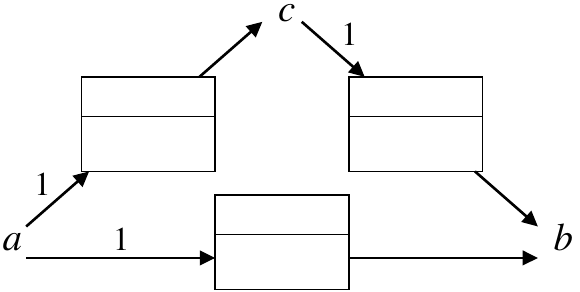
\hfill
\end{center}
\caption{Boolean incoherent feedforward of type 1. 
At right, graphical representation of the \textsc{grn}
$N=(V,M,E_V,E_M,{\cal K})$ with $V=\{a,b,c\}$, the bounds of $a$, $b$
and $c$ are equal to 1, $M=\{l, \lambda, \sigma\}$,
$\phi_{l}$ is $(a\geq 1)$, 
$\phi_{\lambda}$ is $(\lnot(c\geq 1))$, 
$\phi_{\sigma}$ is $(a\geq 1)$. Finally the family of integers is 
$\{K_{a,\emptyset}$,  
$K_{c,\emptyset}$,  
$K_{c,\{l\}}$,  
$K_{b,\emptyset}$,  
$K_{b,\{\sigma\}}$,  
$K_{b,\{\lambda\}}$, 
$K_{b,\{\sigma,\lambda\}}\}$. 
}
\label{figure:FFLtype1}
\end{figure}

\vspace{-\parskip}\label{section:example}
In~\cite{Alon2002-a,Alon2002-b} Uri Alon and co-workers have studied
the most common in vivo patterns involving three genes. Among them,
they have enlightened the ``incoherent feedforward loop of
type~1''. It is composed by a transcription factor $a$ that activates
a second transcription factor $c$, and both $a$ and $c$ regulate a
gene $b$: $a$ is an activator of $b$ whereas $c$ is an inhibitor of
$b$. There is a ``short'' positive action of $a$ on $b$ and a ``long''
negative action \textit{via} $c$: $a$ activates $c$ which inhibits
$b$. The left hand side of Figure~\ref{figure:FFLtype1} shows such a
feedforward loop. Considering that both thresholds of actions of $a$
are equal leads to a boolean network since, in that case, the
variable $a$ can take only the value 0 ($a$ has no action) or 1 ($a$
activates both $b$ and $c$).
The right hand side of the figure shows the corresponding \textsc{grn} with
multiplexes:  $\sigma$ encodes the ``short'' action of $a$ on $b$, whilst $l$
followed by $\lambda$ constitute the ``long'' action.

Several authors, like Uri Alon, consider that if $a$ is equal to 0 for
a sufficiently long time, both $b$ and $c$ will also be equal to 0,
because $b$ and $c$ need $a$ as a resource in order to reach the
state~1. They also consider that the function of this feedforward loop
is to ensure a \emph{transitory activity} of $b$ that signals when $a$
has switched from 0 to 1: the idea is that $a$ activates the
productions of $b$ and $c$, and then $c$ stops the production of $b$.

Here, we take a look at this question \textit{via} four different path
programs, and we prove formally that this affirmation is only valid under some
constraints on the parameters of the network, and only under the assumption
that $b$ starts its activity before $c$.

\paragraph{Is a  transitory production of $b$ possible?}
As already stated, the function classically associated
with the feedforward loop is to ensure a \emph{transitory activity} of
$b$ that signals when $a$ has switched from 0 to 1.  An interesting
question is \emph{under which conditions the previous property is
true~?}
For example the path program 
\begin{eqnarray}
{\cal P}_1 \equiv (b+ ; c+ ; b-)
\end{eqnarray} 
together with the pre-condition $P\equiv (a=1~\wedge ~b=0~\wedge
~c=0)$ and the post-condition $Q_0\equiv \{b=0\}$, is a possible
formalization of the previous property about the behaviour of the
feedforward loop. The backward strategy using our genetically modified
Hoare logic on this example gives the
following successive conditions.
\begin{itemize}\vspace{\itemsep}
	\item\vspace{-\itemsep}
The weakest precondition obtained through the last instruction ``$b-$''
is the following conjunction:
$$
\Phi_b^- \land Q_0[b\leftarrow  b-1]  ~~ \equiv ~~
\left\{\begin{array}{l}
(\neg \neg (c\geq 1) \wedge  \neg (a\geq 1)) \Longrightarrow  K_{b} < b	\\ 
(\neg \neg (c\geq 1) \wedge  (a\geq 1)) \Longrightarrow  K_{b,\sigma } < b\\ 
(\neg (c\geq 1) \wedge  \neg (a\geq 1)) \Longrightarrow  K_{b,\lambda } < b\\ 
(\neg (c\geq 1) \wedge  (a\geq 1)) \Longrightarrow  K_{b,\sigma \lambda } < b\\ 
b - 1 = 0
\end{array}\right.
$$
which simplifies as the conjunction $Q_{1}$:
$$
Q_{1} ~~ \equiv ~~
\left\{\begin{array}{l}
b=1\\ 
((c\geq 1) \wedge  (a<1)) \Longrightarrow  K_{b} = 0\\ 
((c\geq 1) \wedge  (a\geq 1)) \Longrightarrow  K_{b,\sigma } = 0	\\ 
((c<1) \wedge  (a<1)) \Longrightarrow  K_{b,\lambda } = 0	\\ 
((c<1) \wedge  (a\geq 1)) \Longrightarrow  K_{b,\sigma \lambda } = 0
\end{array}\right.
$$
	\item\vspace{-\itemsep}
Then, the weakest precondition obtained through the instruction
``$c+$'' is:
$$
\Phi_c^+ \land Q_1[c\leftarrow c+1]\equiv
\left\{\begin{array}{l}
\neg (a\geq 1) \Longrightarrow  K_{c} > c\\ 
a\geq 1 \Longrightarrow  K_{c,l} > c	\\ 
b=1\\ 
((c+1\geq 1) \wedge  (a<1)) \Longrightarrow  K_{b} = 0\\ 
((c+1\geq 1) \wedge  (a\geq 1)) \Longrightarrow  K_{b,\sigma } = 0\\ 
((c+1<1) \wedge  (a<1)) \Longrightarrow  K_{b,\lambda } = 0\\ 
((c+1<1) \wedge  (a\geq 1)) \Longrightarrow  K_{b,\sigma \lambda } = 0
\end{array}\right.
$$ which simplifies as $Q_{2}$ owing to the boundary axioms and
remarks~\ref{remark:boundaryAxioms}: $$
Q_{2} ~~
\equiv ~~
\left\{\begin{array}{l}
c=0\\ 
a<1 \Longrightarrow  K_{c} = 1	\\ 
a\geq 1 \Longrightarrow  K_{c,l} = 1\\ 
b=1\\ 
a<1 \Longrightarrow  K_{b} = 0	\\ 
a\geq 1 \Longrightarrow  K_{b,\sigma } = 0
\end{array}\right.
$$
	\item\vspace{-\itemsep}
Lastly, the weakest precondition obtained through the first
 ``$b+$'' of the program is:
$$
\Phi_b^+ \land Q_2[b\leftarrow b+1]\equiv
\left\{\begin{array}{l}
(\neg \neg (c\geq 1) \wedge  \neg (a\geq 1)) \Longrightarrow  K_{b} > b	\\ 
(\neg \neg (c\geq 1) \wedge  (a\geq 1)) \Longrightarrow  K_{b,\sigma } > b\\ 
(\neg (c\geq 1) \wedge  \neg (a\geq 1)) \Longrightarrow  K_{b,\lambda } > b\\ 
(\neg (c\geq 1) \wedge  (a\geq 1)) \Longrightarrow  K_{b,\sigma \lambda } > b\\ 
c=0\\ 
a<1 \Longrightarrow  K_{c} = 1	\\ 
a\geq 1 \Longrightarrow  K_{c,l} = 1\\ 
b+1=1\\ 
a<1 \Longrightarrow  K_{b} = 0	\\ 
a\geq 1 \Longrightarrow  K_{b,\sigma } = 0
\end{array}\right.
$$
which simplifies as $Q_{3}$:
$$
Q_{3} ~~ \equiv ~~
\left\{\begin{array}{l}
a<1 \Longrightarrow  K_{b,\lambda } = 1\\ 
a\geq 1 \Longrightarrow  K_{b,\sigma \lambda } = 1	\\ 
c=0\\ 
a<1 \Longrightarrow  K_{c} = 1\\ 
a\geq 1 \Longrightarrow  K_{c,l} = 1\\ 
b=0\\ 
a<1 \Longrightarrow  K_{b} = 0\\ 
a\geq 1 \Longrightarrow  K_{b,\sigma } = 0
\end{array}\right.
$$
\end{itemize}
Then, using the empty program rule, it comes $P\Longrightarrow
Q_3$ \emph{i.e.} $(a=1~\wedge ~b=0~\wedge ~c=0) \Longrightarrow Q_{3}$
and after simplification we get the correctness if and only if
$K_{b,\sigma \lambda } = 1$ and $K_{c,l} = 1$ and $K_{b,\sigma } =
0$. This proves that, whatever the values of the other parameters, the
system can exhibit a transitory production of $b$ in response to a
switch of $a$ from 0 to 1.

\paragraph{Is a transitory production of $b$ possible without
increasing $c$?}  The previous program ${\cal P}_1$ is not the only
one reflecting a transitory production of $b$, there may be other
realisations of this property. For example one can consider the path
program~:
\begin{eqnarray}
{\cal P}_2 \equiv (b+ ; b-).
\end{eqnarray}
With respect to this path program, the weakest precondition obtained through
the last instruction ``$b-$'' is of course $Q_{1}$ as previously.
Then, the weakest precondition obtained through
``$b+$'' is:
$$
Q_{4} ~~ \equiv ~~
\left\{\begin{array}{l}
b=0\\ 
((c\geq 1) \wedge  (a<1)) \Longrightarrow  ((K_{b} = 1) \wedge  (K_{b} = 0))\\ 
((c\geq 1) \wedge  (a\geq 1)) \Longrightarrow  ((K_{b,\sigma }=1) \wedge 
(K_{b,\sigma }=0))\\ 
((c<1) \wedge  (a<1)) \Longrightarrow  ((K_{b,\lambda }=1) \wedge  (K_{b,\lambda }=0))\\ 
((c<1) \wedge  (a\geq 1)) \Longrightarrow  ((K_{b,\sigma \lambda } = 1) \wedge 
(K_{b,\sigma \lambda } = 0))
\end{array}\right.
$$ $Q_{4}$ is of course not satisfiable: it implies that each
parameter associated with $b$ is both equal to 0 and 1. The path
program $(b+ ; b-)$ is not feasible (inconsistent weakest
precondition). Indeed, we retrieve an obvious property of the Thomas'
approach: if $b$ has no negative action on itself then the sequence
$(b+ ; b-)$ cannot arise because the resources of $b$ must change in
order to switch its direction of evolution.

\paragraph{Another possible path compatible with the path program $(b+,c+,b-)$.}
Let us notice that, when $K_{b,\sigma \lambda } = 1$, $K_{c,l} = 1$ and
$K_{b,\sigma } = 0$, even if the system \emph{can} exhibit a transitory
production of $b$ \textit{via} $(b+ ; c+ ; b-)$, this does not prevent from
some other paths that \emph{do not} exhibit this behaviour. For example the
simple path ${\cal P}_3 \equiv c+$ leaves $b$ constantly equal to 0, and
 the Hoare triplet $$
\left\{\begin{array}{l}
a=1~\wedge ~b=0~\wedge ~c=0~\wedge \\ 
K_{b,\sigma \lambda }=1~\wedge ~K_{c,l}=1~\wedge ~K_{b,\sigma }=0
\end{array}\right\}
~ c+ ~
\left\{\begin{array}{l}
b=0
\end{array}\right\}
$$
is satisfied, as the corresponding weakest precondition $Q_{5}$ is clearly
implied by the precondition.
$$
Q_{5} ~~ \equiv ~~
\Phi_c^+ \land Q_0[c\leftarrow c+1] ~~ \equiv ~~
\left\{\begin{array}{l}
c=0\\ 
a=0 \Longrightarrow  K_{c} = 1\\ 
a=1 \Longrightarrow  K_{c,l} = 1	\\ 
b = 0
\end{array}\right.
$$

\paragraph{When $a$ is constantly equal to 1 and when $c=1$, production of $b$ is impossible.}
Even worst: when $a$ is constantly equal to 1, once $c$ has reached
the level 1, it is impossible for $b$ to increase to 1. We prove this
property by showing that the following triplet is inconsistent, whatever the
loop invariant $I$:
\begin{eqnarray}
~~~~~~\left\{\begin{array}{l}
a=1~\wedge ~b=0~\wedge ~c=1~\wedge \\ 
K_{b,\sigma \lambda }\!=\!1~\wedge ~K_{c,l}\!=\!1~\wedge ~K_{b,\sigma }\!=\!0
\end{array}\right\}
\underbrace{
while~b\!<\!1~with~I~do~\exists (b+,b-,c+,c-)}_{{\cal P}_4}
\left\{\begin{array}{l}
b\!=\!1
\end{array}\right\}
\label{eqn:triplet-while1}
\end{eqnarray}
The subprogram $\exists (b+,b-,c+,c-)$ reflects the fact that $a$
stays constant but $b$ or $c$ evolves. The $while$ statement
allows $b$ and $c$ to evolve freely until $b$ becomes equal to 1.
\par\noindent 
Applying the Iteration rule, $I$ has to satisfy:
\begin{itemize}\vspace{\itemsep}
	\item\vspace{-\itemsep}
$\neg (b < 1) \wedge  I \Longrightarrow  (b=1)$ \\  This property is trivially
satisfied whatever the assertion $I$, due to the boundary axioms.
	\item\vspace{-\itemsep}
$\hoare{b < 1\wedge I}{\exists (b+,b-,c+,c-)}{I}$
\\ 
We apply the existential quantifier rule, which gives the following
weakest precondition:
$$
Q_{6}~~\equiv ~~(\Phi^{+}_{b}\wedge \msubst{I}{b}{b+1})~\vee ~
(\Phi^{-}_{b}\wedge \msubst{I}{b}{b-1})~\vee ~
(\Phi^{+}_{c}\wedge \msubst{I}{c}{c+1})~\vee ~
(\Phi^{-}_{c}\wedge \msubst{I}{c}{c-1})
$$
Consequently $I$ can be any assertion such that
\begin{center}
$(b=0\wedge I)\Longrightarrow Q_{6}$
\end{center}
\end{itemize}
Let us denote $P$ the precondition of the path program ${\cal P}_4$.
Applying the empty program rule, it comes that $I$ must also satisfy
$P\Longrightarrow I$. So, because $P\Longrightarrow (b=0)$, we have $P \Longrightarrow  (b=0\wedge I)$,
which, in turn implies $Q_{6}$. Moreover, let us remark that
$$
Q_{6} ~\Longrightarrow ~ (\Phi^{+}_{b}\vee \Phi^{-}_{b}\vee 
\Phi^{+}_{c}\vee \Phi^{-}_{c})
$$
Consequently, if the Hoare triple \ref{eqn:triplet-while1} is correct, then:
$$ P ~ \Longrightarrow  ~ (\Phi^{+}_{b}\vee \Phi^{-}_{b}\vee 
\Phi^{+}_{c}\vee \Phi^{-}_{c})
$$
which is impossible because, if $P$ is satisfied, then:
\begin{itemize}\vspace{\itemsep}
	\item\vspace{-\itemsep}
$\Phi^{+}_{b}$ is false, as $a=1$, $c=1$ and
$K_{b,\sigma }=0$ (indeed,$\Phi^{+}_{b}$ implies $a=1 \land c=1 \Rightarrow
K_{b,\sigma}>0$) 
	\item\vspace{-\itemsep}
$\Phi^{-}_{b}$ is false, as $b=0$ ($\Phi^{-}_{b}$ implies $b>0$)
	\item\vspace{-\itemsep}
$\Phi^{+}_{c}$ is false, as $c=1$ ($\Phi^{+}_{c}$ implies $c<1$)
	\item\vspace{-\itemsep}
$\Phi^{-}_{c}$ is false, as $a=1$, $c=1$ and
$K_{c,l}=1$ 
($\Phi^{-}_{c}$ implies $a=1 \land c=1 \Rightarrow K_{c,l}<1$).  
\end{itemize}
So, we have formally proved that when $a$ is constantly equal to 1, once
$c$ has reached the level 1, it is impossible for $b$ to increase
to~1.
\vspace{-\parskip}\section{
	Partial Correctness and Completeness
}\vspace{-\parskip}\label{WPcorrectness}

"Partial" has to be understood here as "Assuming that the \emph{while}
loops terminate", as usual in Hoare logic.

\subsection{Correctness}

Correctness of our modified Hoare logic means that: if ~ $\vdash
\hoare{P}{p}{Q}$ according to the inference rules of
Section~\ref{discreteHoare}, then the Hoare triple ~ $\hoare{P}{p}{Q}$ ~ is
semantically satisfied according to Definition~\ref{HOAREsemantics}, i.e.: for
all $\eta\in S$ ($S$ being the set of states of $\mathcal{S}$) such that $\eta\models_{N} P$ there exists
$E\subset S$ such that $\eta\relation{p}E$ and $\forall\eta'\in E,
~\eta'\models_{N} Q$.

The proof is made as usual by induction on the proof tree of ~ $\vdash
\hoare{P}{p}{Q}$. Hence, we have to prove that each rule of
Section~\ref{discreteHoare} is correct. Here we develop only the
\emph{Incrementation rule} and the \emph{Sequential composition rule} since
the correctness of the other inference rules is either similar
(\emph{Decrementation rule}) or trivial (\emph{Assert rule}, \emph{quantifier
rules}, \emph{Assignment rule} and \emph{Empty program rule}) or standard in
Hoare Logic (\emph{Alternative rule} and \emph{Iteration rule}). Let us note
that the correctness of the \emph{Sequential composition rule} is neither
trivial nor standard because its semantics is enriched to cope with the
quantifiers. 

Let $N$ be a \textsc{grn} and let $\eta$ be any state of the associated state
space $S$:

\begin{description}
\item[Incrementation rule:]
~ \regle{}{\hoare{~
	\Phi_{v}^{+}~\wedge ~\msubst{Q}{v}{v+1}
	~}{v+}{Q}}
~ (where $v$ is a variable of the \textsc{grn})

From Definition~\ref{HOAREsemantics}, the hypothesis is:

\begin{itemize}
	\item[\fbox{$H$}]
$\eta\models_{N}\Phi_{v}^{+}$ ~ and ~
$\eta\models_{N}\msubst{Q}{v}{v+1}$	
\end{itemize}
and we have to prove the conclusion:
\begin{itemize}
	\item[\fbox{$C$}]
there exists $E\subset S$ such that $\eta\relation{v+}E$
and $\forall\eta'\in E,~\eta'\models_{N} Q$
\end{itemize}

Let us choose $E=\{\eta'\}$ with $\eta'=\msubst{\eta}{v}{\eta(v)+1}$. From
Notation~\ref{lesPHI}, the hypotesis $\eta\models_{N}\phi_{v}^{+}$ is
equivalent to $(\eta\rightarrow\eta')\in\mathcal{S}$, which in turn,
according to Definition~\ref{PATHsemantics}, implies
$\eta\relation{v+}\{\eta'\}$.
Hence, it only remains to prove that $\eta'\models_{N}Q$, which results from
the hypothesis $\eta\models_{N}\msubst{Q}{v}{v+1}$.~\cqfd
\item[Sequential composition rule:]
~ \regle{\hoare{P_2}{p_2}{Q}~~~~~\hoare{P_1}{p_1}{P_2}}
	{\hoare{P_1}{p_1;p_2}{Q}}

From Definition~\ref{HOAREsemantics}, we consider the following three
hypotheses:
\begin{itemize}
	\item[\fbox{$H_1$}]
for all $\eta_1\in S$ such that $\eta_1\models_N P_1$ there exists
$E_1$ such that $\eta_1\relation{p_1}E_1$ and $\forall\eta'\in
E_1,~\eta'\models_N P_2$
	\item[\fbox{$H_2$}]
for all $\eta_2\in S$ such that $\eta_2\models_N P_2$ there exists
$E_2$ such that $\eta_2\relation{p_2}E_2$ and $\forall\eta''\in
E_2,~\eta''\models_N Q$
	\item[\fbox{$H_3$}]
$\eta\models_N P_1$ 
\end{itemize}
and we have to prove the conclusion:
\begin{itemize}
	\item[\fbox{$C$}]
there exists $E\subset S$ such that $\eta\relation{p_1;p_2}E$ and
$\forall\eta''\in E,~\eta''\models_N Q$
\end{itemize}

Let us arbitrarily choose a set $E_1$ such that $\eta\relation{p_1}E_1$ and
$\forall\eta'\in E_1,~\eta'\models_N P_2$ (we know that $E_1$ exists from
\fbox{$H_1$} and \fbox{$H_3$}).

For each $\eta'\in E_1$, we similarly choose a set $E_2^{\eta'}$ such
that:
\\

$\eta'\relation{p_2}E_2^{\eta'}$ and $\forall\eta''\in
E_2^{\eta'},~\eta''\models_N Q$ (we know that the family
$\{E_2^{\eta'}\}_{\eta'\in E_1}$ exists from \fbox{$H_2$} and the fact that
$\eta'\models_N P_2$ for all $\eta'\in E_2$)

Let $E=(\bigcup_{\eta'\in E_1}E_2^{\eta'})$, we have:
$\eta\relation{p_1;p_2}E$ from Definition~\ref{PATHsemantics} and
$\forall\eta''\in E,~\eta''\models_N Q$ (from the way the union is built).
\cqfd
\end{description}
\subsection{Weakest precondition}

Completeness of Hoare logic would be of course defined as follows: If the
Hoare triple $\hoare{P}{p}{Q}$ is satisfied (according to
Definition~\ref{HOAREsemantics}) then ~ $\vdash\hoare{P}{p}{Q}$ (using the
inference rules of Section~\ref{discreteHoare} as well as first order logic
and proofs on integers).

Obviously, as such, Hoare logics cannot be complete because, as already
mentioned for classical Hoare logic, finding the weakest loop invariants is
undecidable and there is no complete logic on integers (G\"odel). So,
following Dijkstra~\cite{Dijkstra:1975:GCN:360933.360975}, we prove
completeness under the assumptions that the loop invariants of all $while$
statements are weakest invariants and that the needed properties on
integers are admitted. We adopt the strategy that computes the weakest
precondition and we will prove the following theorem:

\begin{theorem}\emph{(Dijkstra theorem on the genetically modified
Hoare logic)} 
A \textsc{grn} $N$ and a Hoare triple $\hoare{P}{p}{Q}$ being given, the
backward strategy defined at the end of Section~\ref{stdHoare}, with the inference
rules of Section~\ref{discreteHoare}, computes the \emph{weakest precondition}
$P_0$ just before the last inference that uses the \emph{Empty program rule}.
\\
It means that: if $\hoare{P}{p}{Q}$ is satisfied, then $P\Rightarrow P_0$
is satisfied.
\end{theorem}

This theorem has an obvious corrolary:
\begin{corollary}
A \textsc{grn} $N$ being given, our modified Hoare logic is complete under the
assumption that all given loop invariants are the weakest ones and that the
needed properties on integers are established.
\end{corollary}
\textbf{Proof of the corollary:}
if $\hoare{P}{p}{Q}$ is satisfied, then, from the Dijkstra theorem above,
there is a proof tree that infers the Hoare triple if there is a proof tree
for the property $P\Rightarrow P_0$ (which is semantically satisfied because
$P_0$ is the weakest precondition). First order logic being complete and
properties on integers being axiomatically assumed, the proof tree for
$P\Rightarrow P_0$ exists.\cqfd

\paragraph{Proof of Dijkstra theorem:}\mbox{}

\noindent
Under the hypotheses that all loop invariants are minimal and that the Hoare
triple $\hoare{P}{p}{Q}$ is satisfied, i.e., under the hypotheses:
\begin{itemize}
	\item[\fbox{$H_1$}]
for all $\eta$ satisfying $P$, there exists $E$ such that $\eta\relation{p}E$
and for all $\eta'\in E$, $\eta'$ satisfies $Q$
	\item[\fbox{$H_2$}]
for all $while$ statements of $p$, the corresponding loop invariant $I$ is the
weakest one
\end{itemize}
one has to prove the conclusion:
\begin{itemize}
	\item[\fbox{$C$}]
$P\Rightarrow P_0$ is satisfied, where $P_0$ is the precondition computed by
the proof of $\hoare{P}{p}{Q}$ according to the backward strategy with the inference
rules of Section~\ref{discreteHoare}.
\end{itemize}
The proof is done by structural induction according to the backward strategy on $p$.
\begin{itemize}
	\item
If $p$ is of the form $while~e~with~I~do~p'$, then, by construction of the
backward strategy, applying the \emph{Iteration rule}, we get $P_0=I$, and the
conclusion results immediately from \fbox{$H_2$}.
	\item
If $p$ is of the form $v+$, then the only set $E$ such that
$\eta\relation{v+}E$ is $E=\{\msubst{\eta}{v}{v+1}\}$. The hypothesis
\fbox{$H_1$} becomes:
	\begin{itemize}
		\item[\fbox{$H_1$}]
	for all $\eta$ satisfying $P$, $\eta'=\msubst{\eta}{v}{v+1}$
	satisfies~$Q$ and $\eta\rightarrow\eta'$ is a transition
	of~$\mathcal{S}$
	\end{itemize}
and from the \emph{Incrementation rule}, the conclusion becomes:
	\begin{itemize}
		\item[\fbox{$C$}]
	$P\Rightarrow (\Phi_v^{+}\wedge\msubst{Q}{v}{v+1})$ is satisfied.
\end{itemize}
So, \fbox{$H_1$} $\Rightarrow$ \fbox{$C$} straightforwardly results from the
definition of $\Phi_{v^{+}}$ (Notation~\ref{lesPHI}) and we do not use
\fbox{$H_2$}.
	\item
If $p$ is of the form $p_1;p_2$, then we firstly inherit the two structural
induction hypotheses:
	\begin{itemize}
		\item[\fbox{$H_3$}]
	for all assertions $P'$ and $Q'$, if $\hoare{P'}{p_1}{Q'}$ is
	satisfied then $P'\Rightarrow P_1$ is satisfied, where $P_1$ is the
	precondition computed from $Q'$ \textit{via} the backward strategy
		\item[\fbox{$H_4$}]
	for all assertions $P''$ and $Q''$, if $\hoare{P''}{p_2}{Q''}$ is
	satisfied then $P''\Rightarrow P_2$ is satisfied, where $P_2$ is the
	precondition computed from $Q''$ \textit{via} the backward strategy
	\end{itemize}
Moreover the hypothesis \fbox{$H_1$} becomes (Definition~\ref{PATHsemantics}):
	\begin{itemize}
		\item[\fbox{$H_1$}]
	for all $\eta$ satisfying $P$, there exists a family of state sets
	$\mathcal{F}=\{E_e\}_{e\in F}$ such that $\eta\relation{p_1}F$ and
	$e \relation{p_{2}} E_{e}$ for all $e\in F$ and for all $\eta'\in
	E=(\bigcup_{e\in F}E_{e})$, $\eta'$ satisfies $Q$
	\end{itemize}
Lastly, from the \emph{Sequential composition rule}, the conclusion becomes:
	\begin{itemize}
		\item[\fbox{$C$}]
	$P\Rightarrow P_1$ is satisfied, where $P_1$ is the weakest
	precondition of $\hoare{\cdots}{p_1}{P_2}$, $P_2$ being the weakest
	precondition of $\hoare{\cdots}{p_2}{Q}$.
	\end{itemize}
From \fbox{$H_4$} (with $Q''=Q$) it results that all the states $e\in F$ of
hypothesis \fbox{$H_1$} satisfy $P_2$. Consequently $\hoare{P}{p_1}{P_2}$ is
satisfied. Thus, from \fbox{$H_3$} (with $Q'=P_2$ and $P'=P$) it comes
$P\Rightarrow P_1$, which proves the conclusion.
	\item
Similarly to the correctness proof, we do not develop here the other cases of
the structural induction. They are either similar to already developed cases
(\emph{Decrementation rule}) or trivial (\emph{Assert rule}, \emph{quantifier
rules}, and \emph{Assignment rule}) or standard in Hoare Logic
(\emph{Alternative rule}).
\end{itemize}
This ends the proof.\cqfd
\vspace{-\parskip}\section{
	Discussion
}\label{discussion}\vspace{-\parskip}

The cornerstone of the modeling process lies, whatever the application
domain, in the determination of parameters. In this paper, we proposed
an approach for exhibiting constraints on parameters of gene network
models, that relies on the adaptation of the Hoare logic, initially
designed for proofs of imperative programs. It leads to several
questions about its usability and implementations.

\subsection{Language issues}

The path language is a way to describe formally the specification of
correct models of gene networks. Classically, the specifications can
be expressed in temporal logics, like CTL and LTL, which also allows
the modeler to take into account behavioral information. But even if
there exists some links between path language and temporal logics,
these formal languages (temporal logics and path language) are not
comparable: some properties can be expressed in the path language and
not in classical temporal logics and converselly.
\begin{itemize}
\item On the one hand, in the path program, the assignment instruction
allows one to express a knock-out of a gene ($v:=0$). Such knock-out
of a gene is not expressible in CTL or LTL.
\item On the other hand, CTL or LTL is able to express properties on
infinite cyclic traces. Such properties on infinite traces would be
expressed in the path language by a program which does not terminate,
and consequently, the post-condition would not make sense. 
\end{itemize}
Nevertheless, a succession of incrementation/decrementation
instructions corresponds to a property that can be expressed in the CTL
language. For example, if one knows the
starting point, say $(v_1=1 \land v_2=0)$, the path
program $v_1+;v_2+;v_1-$ corresponds to the formula
$EX(v_1=2 \land EX(v_2=1 \land EX(v_1=1 \land v_2=0)))$. 
Correctness of this program path with the precondition $(v_1=1 \land
v_2=0)$ becomes equivalent to verify that the CTL formula 
$(v_1=1 \land
v_2=0) \Rightarrow EX(v_1=2 \land EX(v_2=1 \land EX(v_1=1 \land
v_2=0)))$ is true in all possible states. 
More generally, the path language is well suited for
sequential properties whereas CTL can express non sequential ones.

In the path language, invariants of $while$ loops are
mandatory: Hoare logic is able to prove a program with
$while$ statements only if invariants are given. In other words, the
entire information that the Hoare logic needs to perform the proof, is
given by invariants. Unfortunately, invariants are difficult to
devise. Thus the $while$ statements are often used in proofs by
refutation, where the proof is done for each possible invariant, see our
example of section~\ref{example}.

\subsection{Plateform issues}
The Hoare logic for gene networks has been designed in order to
support a software which aims at helping the determination
of parameters of models of gene networks. We have already done its
proof of feasability. Indeed, after having developped a prototype
named \texttt{SMBioNet} which enumerates all possible valuations of
parameters and retains only those which are coherent with a specified
temporal logic formula, we developped a new prototype
called \texttt{WP-SMBioNet}~\cite{ZohraKhalisPhD},
which uses a path program and the Weakest Precondition calculus
(backward strategy) to 
produce constraints on parameters.

In order to compare both approaches (CTL formulae \emph{versus} 
path programs), we consider a property which can be expressed in both
temporal logic CTL and path program. When modeling the biological
system triggering the tail resorption during the metamorphose of
tadpole, see~\cite{articleJBPC2007} and references therein, the
expression profiles of \cite{Leloup1977} can be translated into a path
program, see Fig~\ref{figure:comparaisonCTLHoare-1}, which in turn can
be translated into an equivalent CTL formula. For this example,
whereas \texttt{SMBioNet} needs more than 3 hours to selects among all
possible parameterizations those which lead to a dynamics coherent with
the CTL formula, \texttt{WP-SMBioNet} needs only 10 seconds (on the
same computer) to construct the constraints on the parameters. If we
ask the enumeration of the parameters satisfying the constraints
(using Choco~\cite{choco-ref-RR-EMN}), the total search time is
about 2 minutes. This example shows that the Hoare logic can speed up
the determination of coherent parameterizations.

\begin{figure}
\begin{center}
\begin{minipage}{14cm}
\begin{verbatim}
# initializing 
T3:=1; T4:=1; d3:=1; d2:=0; gi:=0; gp:=0; gt:=0; tr:=0;
# evolutions 
gi+; d2+; T3+; tr+; T3+; gp+; d3-; gt+;
\end{verbatim}
\end{minipage}
\end{center}

$$
\left(
\begin{array}{ll}
(T3=1) & \land \\
(T4=1) & \land \\
(d3=1) & \land \\
(d2=0) & \land \\
(gi=0) & \land \\
(gp=0) & \land \\
(gt=0) & \land \\
(tr=0)
\end{array}
\right)
\Rightarrow
\left(
\begin{array}{l}
EX((gi=1) \land \\
\hspace*{9mm}   EX((d2=1) \land \\
\hspace*{18mm}      EX((T3=2) \land \\
\hspace*{27mm}         EX((tr=1) \land \\
\hspace*{36mm}            EX((T3=3) \land \\
\hspace*{45mm}               EX((gp=1) \land \\
\hspace*{54mm}                  EX((d3=0) \land  \\
\hspace*{63mm}                     EX(gt=1))))))))
\end{array}
\right)
$$
\caption{A path program (top) with its corresponding CTL formula (bottom)}
\label{figure:comparaisonCTLHoare-1}
\end{figure}

We can notice that the complexity of the weakest precondition calculus
is linear with the number of instructions in the path program, and does not
depend on the size of gene regulatory networks: each node of
the syntaxic tree of the program is visited only once. At the opposite,
the CTL model checking algorithm depends on the size of the
network. Thus, the use of path program instead of CTL formula leads to
postpone the enumeration step which then can use the constraints on
parameters to cut down drastically the set of parameterizations to
consider.

A software plateform dedicated to analysis of gene regulatory
networks, should have to combine different technics. Indeed constraints
solving technics are necessary to enumerate parameters or give
counter-examples, theorem prover can be also useful to get strategies
for proofs by refutation, and model checking technics and Hoare logic
precondition calculus should be combined in order to give very
efficient algorithm. As already noted, it seems natural to use Hoare logic when
the behavioural specification focuses on a finite time horizon,
whereas model checking is natural when the temporal specification
expresses global properties on infinite traces.

It would be interesting to complete this plateform with some improved
features. From a theoretical point of view, one could also develop
approaches to help finding loop invariants. To build them, it seems
possible to adapt the iterative approach adopted in
ASTREE~\cite{Cousot2005a} but in another context (abstract
interpretation~\cite{Cousot2004a}): Pragmatically one begins with a
simple invariant $I$, then one tries to make the proof and completes
iterativelly and partially the invariant. From an application point of
view, specifications often stem from DNA profiles, it would be
valuable to develop a program that automatically produces path
programs from DNA chips data. Two questions emerge: the choice of
thresholds on which is based the discretization of expression levels,
and the determination of good time steps. 
These questions are out of the scope of this article. They mainly rely
on biological expertise and experimental conditions.

\vspace{-\parskip}\section*{
	Acknowledgment
}\vspace{-\parskip}

We are grateful to Alexander Bockmayr and Heike Siebert for fruitful
discussions and comments on this paper. The authors thank the
French National Agency for Research (ANR-10-BLANC-0218 BioTempo
project) for its support. This work has also been  supported
by the CNRS PEPII project "\emph{CirClock}".
\bibliographystyle{alpha}
\bibliography{biblioHoareLogic,biblio-JPC}

\newcommand{\etalchar}[1]{$^{#1}$}
\begin{thebibliography}{SOMMA02}

\bibitem[BCRG04]{articleJTB2004}
G.~Bernot, J.-P. Comet, A.~Richard, and J.~Guespin.
\newblock Application of formal methods to biological regulatory networks:
  Extending {T}homas' asynchronous logical approach with temporal logic.
\newblock {\em Journal of Theoretical Biology}, 229(3):339--347, 2004.

\bibitem[BG01]{Blass2001}
A.~Blass and Y.~Gurevich.
\newblock Inadequacy of computable loop invariants.
\newblock {\em ACM Transactions on Computational Logic}, 2(1), 2001.

\bibitem[BPGT01]{boileau01}
C.~Boileau, J.~Prados, J.~Geiselmann, and L.~Trilling.
\newblock Using constraint programming for learning from experiments
  transcriptional activation and the geometry of {DNA}.
\newblock In T.~Schiex L.~Duret, C.~Gaspin, editor, {\em Actes des Journées
  Ouvertes Biologie Informatique Mathématiques (JOBIM), Toulouse}, 2001.

\bibitem[CC04]{Cousot2004a}
P.~Cousot and R.~Cousot.
\newblock {\em Building the Information Society}, chapter Basic Concepts of
  Abstract Interpretation., pages 359--366.
\newblock Kluwer Academic Publishers, 2004.

\bibitem[CCF{\etalchar{+}}05]{Cousot2005a}
P.~Cousot, R.~Cousot, J.~Feret, L.~Mauborgne, A.~Miné, D.~Monniaux, and
  X~Rival.
\newblock The {ASTRÉE} analyser.
\newblock In M.~Sagiv, editor, {\em ESOP 2005 The European Symposium on
  Programming}, number 3444 in LNCS, pages 21--30. Springer, 2005.

\bibitem[{C}or08]{CORBLIN:2008:TEL-00388776:1}
{F}abien {C}orblin.
\newblock {\em {C}onception et mise en oeuvre d'un outil d{\'e}claratif pour
  l'analyse des r{\'e}seaux g{\'e}n{\'e}tiques discrets}.
\newblock PhD thesis, {U}niversit{\'e} {J}oseph-{F}ourier - {G}renoble {I}, 12
  2008.

\bibitem[cT10]{choco-ref-RR-EMN}
choco Team.
\newblock choco: an open source java constraint programming library.
\newblock Research report 10-02-INFO, Ecole des Mines de Nantes, 2010.

\bibitem[CTF{\etalchar{+}}09]{Corblin2009}
F.~Corblin, S.~Tripodi, E.~Fanchon, D.~Ropers, and L.~Trilling.
\newblock A declarative constraint-based method for analyzing discrete genetic
  regulatory networks.
\newblock {\em Biosystems}, 98(2):91--104, 2009.

\bibitem[Dij75]{Dijkstra:1975:GCN:360933.360975}
Edsger~W. Dijkstra.
\newblock Guarded commands, nondeterminacy and formal derivation of programs.
\newblock {\em Commun. ACM}, 18:453--457, August 1975.

\bibitem[FCT{\etalchar{+}}04]{FanchonCTHG04}
E.~Fanchon, F.~Corblin, L.~Trilling, B.~Hermant, and D.~Gulino.
\newblock Modeling the molecular network controlling adhesion between human
  endothelial cells: Inference and simulation using constraint logic
  programming.
\newblock In {\em CMSB}, pages 104--118, 2004.

\bibitem[Hat74]{HATCHER1974}
W.S. Hatcher.
\newblock A semantic basis for program verification.
\newblock {\em J. of Cybernetics}, 4(1):61--69, 1974.

\bibitem[Hoa69]{logique-Hoare-1969}
C.A.R. Hoare.
\newblock An axiomatic basis for computer programming.
\newblock {\em Communications of the ACM}, 12(10):576--585, oct 1969.

\bibitem[KCRB09]{GGG2009}
Z.~Khalis, J.-P. Comet, A.~Richard, and G.~Bernot.
\newblock The {SMBioNet} method for discovering models of gene regulatory
  networks.
\newblock {\em Genes, Genomes and Genomics}, 3(special issue 1):15--22, 2009.

\bibitem[Kha10]{ZohraKhalisPhD}
Z.~Khalis.
\newblock {\em Logique de Hoare et identification formelle des paramètres d'un
  réseau génétique}.
\newblock PhD thesis, University of Evry-Val d'Essonne, 2010.

\bibitem[LB77]{Leloup1977}
J.~Leloup and M.~Buscaglia.
\newblock La triiodothyronine, hormone de la métamorphose des amphibiens.
\newblock {\em CR Acad. Sci.}, 284:2261--2263, 1977.

\bibitem[MGCLG07]{articleJBCB2007}
D.~Mateus, J.-P. Gallois, J.-P. Comet, and P.~Le~Gall.
\newblock Symbolic modeling of genetic regulatory networks.
\newblock {\em Journal of Bioinformatics and Computational Biology},
  5(2B):627--640, 2007.

\bibitem[MSOI{\etalchar{+}}02]{Alon2002-b}
R.~Milo, S.~Shen-Orr, S.~Itzkovitz, N.~Kashtan, D.~Chklovskii, and U.~Alon.
\newblock Network motifs: Simple building blocks of complex networks.
\newblock {\em Science}, 298:824--827, 2002.

\bibitem[SOMMA02]{Alon2002-a}
S.~Shen-Orr, R.~Milo, S.~Mangan, and U.~Alon.
\newblock Network motifs in the transcriptional regulation network of
  escherichia coli.
\newblock {\em Nature Genetics}, 31:64--68, 2002.

\bibitem[Td90]{TA90}
{R.} Thomas and {R.} {d'}Ari.
\newblock {\em Biological Feedback}.
\newblock CRC Press, 1990.

\bibitem[Tho91]{TH091}
{R.} Thomas.
\newblock Regulatory networks seen as asynchronous automata : A logical
  description.
\newblock {\em J. theor. Biol.}, 153:1--23, 1991.

\bibitem[TK01]{TK2001-b}
R.~Thomas and M.~Kaufman.
\newblock Multistationarity, the basis of cell differentiation and memory.
  {II.} logical analysis of regulatory networks in terms of feedback circuits.
\newblock {\em Chaos}, 11:180--195, 2001.

\bibitem[TTB{\etalchar{+}}07]{articleJBPC2007}
S.~Troncale, R.~Thuret, C.~Ben, N.~Pollet, J.-P. Comet, and G.~Bernot.
\newblock Modelling of the {TH}-dependent regulation of tadpole tail
  resorption.
\newblock {\em Journal of Biological Physics and Chemistry}, 7(2):45--50, 2007.

\end{thebibliography}
\end{document}